\documentclass[footinbib,a4paper,aps,prl,superscriptaddress,reprint,twocolumn,preprintnumbers,amsmath,amssymb,nobalancelastpage,10pt]{revtex4-1}
\usepackage{amsmath}
\usepackage{verbatim}
\usepackage{graphicx}
\usepackage{color}
\usepackage{tikz}
\usepackage{physics}
\usepackage{subfigure}

\usepackage{float}

\usepackage{mathptmx}
\usepackage{amssymb}
\usepackage{amsmath}
\usepackage{amsfonts}
\usepackage{array}

\usepackage{bm}
\usepackage{xcolor}
\usepackage{graphicx}

\usepackage{braket}

\definecolor{mygrey}{gray}{0.35}
\definecolor{myblue}{rgb}{0.2,0.2,0.8}
\definecolor{myzard}{cmyk}{0,0,0.05,0}
\definecolor{mywhite}{rgb}{1,1,1}
\definecolor{myred}{rgb}{1,0.,0.3}

\usepackage[colorlinks=true,citecolor=myblue,linkcolor=myred]{hyperref}

 \def\ee{\mathord{\rm e}}
 
 \def\ii{\mathord{\rm i}}

\renewcommand{\ii}{{\rm i}}
\renewcommand{\ee}{{\rm e}}
\def\beq{\begin{equation}}
\def\eeq{\end{equation}}

\def\barray{\begin{eqnarray}}
\def\earray{\end{eqnarray}}

\begin{document}


\title{Intertwined Topological Phases induced by Emergent Symmetry Protection}

\author{D. Gonz\'{a}lez-Cuadra}\email{daniel.gonzalez@icfo.eu}
\affiliation{ICFO-Institut de Ci\`encies Fot\`oniques, Av. Carl Friedrich Gauss 3, 08860 Barcelona, Spain}

\author{A. Bermudez}
\affiliation{Departamento de F\'{i}sica Te\'{o}rica, Universidad Complutense, 28040 Madrid, Spain}

\author{P. R. Grzybowski}
\affiliation{ICFO-Institut de Ci\`encies Fot\`oniques, Av. Carl Friedrich Gauss 3, 08860 Barcelona, Spain}
\affiliation{Faculty of Physics, Adam Mickiewicz University, Umultowska 85, 61-614 Pozna{\'n}, Poland}
\author{M. Lewenstein}
\affiliation{ICFO-Institut de Ci\`encies Fot\`oniques, Av. Carl Friedrich Gauss 3, 08860 Barcelona, Spain}
\affiliation{ICREA-Instituci\'o Catalana de Recerca i Estudis Avan\c cats, Lluis Companys 23, 08010 Barcelona, Spain}

\author{A. Dauphin}\email{alexandre.dauphin@icfo.eu}
\affiliation{ICFO-Institut de Ci\`encies Fot\`oniques, Av. Carl Friedrich Gauss 3, 08860 Barcelona, Spain}

\begin{abstract}

The dual role played by symmetry in many-body physics manifests itself through two fundamental mechanisms: spontaneous symmetry breaking and topological symmetry protection. These two concepts, ubiquitous in both condensed matter and high energy physics, have been applied successfully in the last decades to unravel a plethora of complex phenomena. Their interplay, however, remains largely unexplored. Here we report how, in the presence of strong correlations, symmetry protection emerges dynamically from a set of configurations enforced by another broken symmetry. This novel mechanism spawns different intertwined topological phases, where topological properties coexist with long-range order. Such a singular interplay gives rise to interesting static and dynamical effects, including interaction-induced topological phase transitions constrained by symmetry breaking, as well as a self-adjusted fractional pumping. This work paves the way for further exploration of exotic topological features in strongly-correlated quantum systems.

\end{abstract}

\maketitle

\paragraph{Introduction.--} The notion of symmetry is paramount to unveil the fundamental laws  of   Nature, while  spontaneous symmetry breaking (SSB)  is essential to understand Nature's   different guises~\cite{Gross14256}. In particular,  at long length-scales, various phases of matter can be understood by the pattern of SSB and  corresponding local order parameters~\cite{landau_symm_breaking}. Although different SSB patterns tend to compete with one another,  a genuine  cooperation  can also arise in strongly-correlated systems with intertwined orders~\cite{RevModPhys.87.457}. More recently, topology has been recognised  as an exotic  driving force  shaping   the texture of Nature, and leading to  phases characterised by topological invariants rather than by local order parameters~\cite{review_wen_top_phases_matter}. It is no longer the breaking of  certain symmetries but, actually, their conservation~\cite{classification_review}, which  gives rise to novel states of matter, the so-called   symmetry-protected topological (SPT) phases~\cite{spt_phases_review}.  In the non-interacting limit, topological insulators and superconductors provide well-understood  examples of this   paradigm~\cite{ti_review_2}, while current  research   aims at understanding strong-correlation effects, such as the  competition of SPT and  SSB phases due to  interactions~\cite{review_int_top_ins_II}.

Alternatively, a cooperation between SPT and SSB may allow for {\it intertwined topological phases} that  simultaneously display  a local order parameter and a topological invariant. For integer and fractional Chern insulators, such intertwined orders have  been already identified in the literature~\cite{int_QAH_honeycomb_MF,interaction_induced_top_insulators, SBT_fermions_triang_lattice}. Nonetheless, in these cases, the topological phases exist in the absence of any protecting symmetry. In more generic situations, the existence of intertwined topological phases  will depend on  how the symmetry responsible for the SPT phase can be embedded into the broader  symmetry-breaking phenomenon. Arguably, the first instance of this situation is the Peierls instability~\cite{peierls} in polyacetylene,  neatly accounted for via the Su-Schrieffer-Heeger (SSH) model at half-filling~\cite{ssh_polymers}. Here, the instability leads to a dimerized lattice distortion and a  bond-order-wave (BOW), where electrons are distributed in an alternating sequence of bonding and anti-bonding orbitals. A closer inspection shows that inversion symmetry is automatically fixed by such a SSB  pattern, which leads to a   topological quantization of the electronic polarization~\cite{PhysRevB.83.245132}, and is ultimately responsible for the protection of the SPT phase.

\begin{figure}[H]
\includegraphics[width=1.0\linewidth]{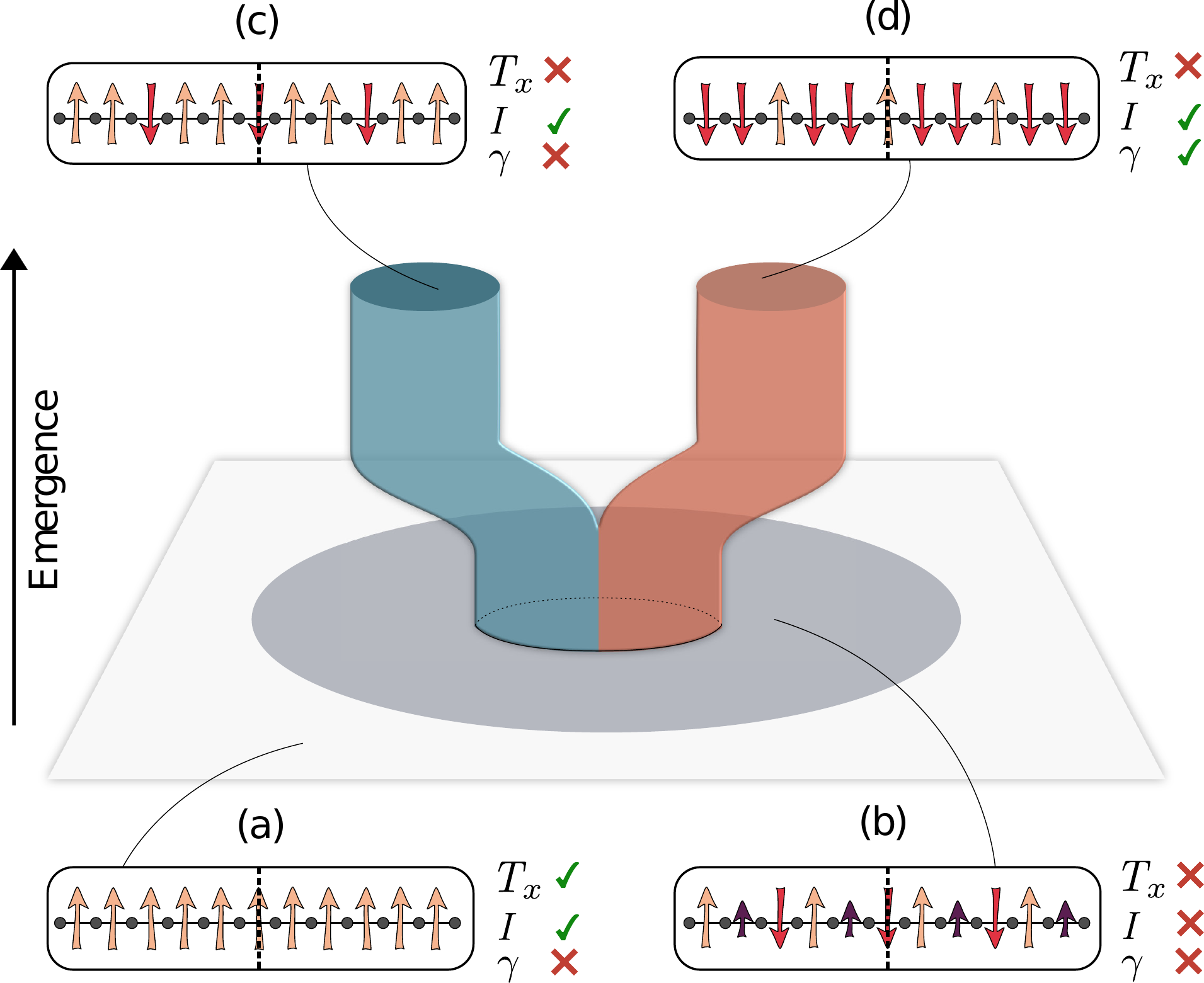}
\caption{\label{fig:emergence}\textbf{Emergent Symmetry Protection:} We represent qualitatively a ground state manifold where different quantum phases are characterized by their symmetry and topological properties, and use   spin patterns to exemplify the different ground-state configurations. (a) Ground state satisfying both translation ($\mathit{T}_x$) and inversion ($\mathit{I}$) symmetry, but lacking any non-zero topological invariant ($\gamma$). The spontaneous breaking of translation symmetry results in a phase with a three-site unit cell, represented in (b) with different arrows accounting for the three possible magnetizations, which may not respect the inversion symmetry, lacking a non-zero topological invariant. Remarkably, such inversion symmetry can  emerge   from all the possible configurations constrained by the SSB pattern, leading to the low-energy sectors depicted in (c, d). Note that these two phases are not only distinguished by the SSB pattern but also, and more importantly,  by  topology. Accordingly, whereas (c) is topologically trivial, (d) presents both a local order parameter and a non-zero topological invariant, and thus corresponds to an intertwined topological phase where the protecting symmetry emerges dynamically.}
\end{figure}

In this work, we study a hitherto unknown possibility:  the emergence of an intertwined topological phase when the SSB pattern does not automatically impose the   protecting  symmetry  (Fig.~\ref{fig:emergence}). Instead, this protecting symmetry {\it emerges} from a larger set of configurations allowed by the SSB, such that  its interplay with  topology and strong correlations  endows the system with very interesting, yet mostly unexplored, static and dynamical behaviour.  We demonstrate  this novel topological mechanism  in  the $\mathbb{Z}_2$-Bose-Hubbard model~\cite{z2bhm_BOW_phases,z2_bhm_SPT_phases}, a microscopic lattice model that displays strongly-correlated intertwined topological phases at various fractional fillings.  At third-filling and for sufficiently-strong interactions, we find a period-3 BOW with a three-fold degenerate ground state  that displays a non-zero topological invariant: the total Berry phase.  We show that  inversion symmetry  emerges from the larger SSB landscape of a bosonic Peierls' mechanism,  protecting the intertwined topological BOW, and making it fundamentally different from other non-topological  BOWs. We unveil a rich phase diagram with first- and second-order quantum phase transitions caused by the interplay of this emergent symmetry, topology and strong correlations.

We also identify a dynamical manifestation of  the underlying topology that is genuinely rooted in strong correlations and the interplay of the emergent and symmetry-broken symmetries: a {\it self-adjusted fractional pump}.  As discussed by Thouless et {\it al}.~\cite{thouless1983,pumping_thouless_int}, the quantization of  adiabatic charge transport in weakly-interacting insulators  uncovers a profound connection to  higher-dimensional topological phases, as recently exploited in cold-atom experiments~\cite{pumping_exp_fermions,pumping_exp_bosons}. Strong interactions can lead to fractional  pumped charges~\cite{pumping_fractional_1,pumping_fractional_2}, showing  a clear reminiscence to the fractional quantum Hall effect (FQHE)~\cite{tao1983,Bergholtz2008,Guo2012,Budich2013}. We show that, following  a  dynamical modulation of the interactions in the $\mathbb{Z}_2$-Bose-Hubbard model, the system self-adjusts within the landscape of SSB sectors, allowing for a cyclic  path that displays a fractional pumped charge~$1/3$,  such that  the correlated intertwined topological phase has no free-particle counterpart. 

\begin{figure}[t]
\includegraphics[width=1.0\linewidth]{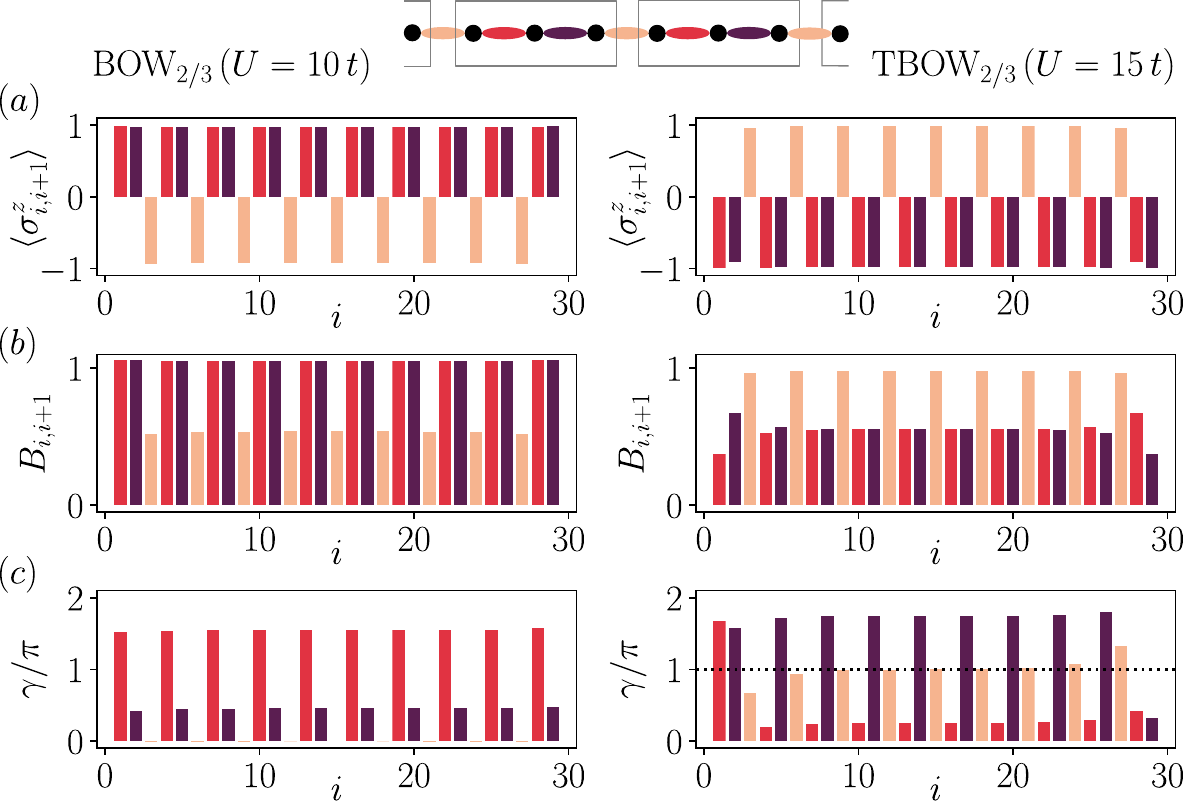}
\caption{\label{fig:berry_trimer}\textbf{Simultaneous orders in intertwined topological phases:}  Real-space configuration of (a) the $\mathbb{Z}_2$ field $\langle \sigma_{i,i+1}^z\rangle$  and (b) the bosonic  bond-densities $B_{i,i+1}=\langle b^\dagger_i b_{i+1}\rangle+{\rm c.c.}$, using different colours for each  element of the unit cell.  Different permutations within the unit cell lead to a 3-fold quasi-degenerate groundstate, each obtained from one another  by translating  the modulation patterns of the ferrimagnetic and BOW orders. The quasi-degeneracy comes from the finite-size effects, but degeneracy is recovered in the thermodynamic limit.  (c)   The Local Berry phases $\gamma$ display  a quantized value  of $0$ ($U=10\,t$) or $\pi$ ($U=15\,t$) on the bonds preserving the inversion symmetry  of the unit cell, allowing us to distinguish between the trivial and topological BOW phases.   We note that  the  topological BOW phase (right panels) does not have a fermionic analogue~\cite{trimer_cylinder} in the groundstate of the   SSH model~\cite{fractionalization_trimer,PhysRevB.27.370}, which instead realizes the trivial BOW  (left panels) for energetic reasons.}
\end{figure}

\paragraph{$\mathbb{Z}_2$-Bose-Hubbard model.--} We consider a 1D system of interacting bosons coupled to a dynamical $\mathbb{Z}_2$ field and described by the lattice Hamiltonian

\begin{equation}
\label{eq:z2bhm}
\begin{aligned}
H =& -\sum_i \left[ b_i^\dagger(t+\alpha\sigma^z_{i,i+1})b_{i+1}^{\phantom{\dagger}}+\text{H.c.}\right] + \frac{U}{2}\sum_i n_i(n_i-1)\\
&+\frac{\Delta}{2}\sum_i\sigma^z_{i,i+1}+\beta\sum_i\sigma^x_{i,i+1},
\end{aligned}
\end{equation}
where $b^\dagger_i$ is the bosonic creation operator at site $i$, $n_i=b^\dagger_i b_i$ is the number operator, and $\sigma^x_{i,i+1},\sigma^z_{i,i+1}$ are the Pauli matrices associated to the $\mathbb{Z}_2$ fields on the bond $(i,i+1)$. The bare bosonic Hamiltonian depends on the hopping strength $t$, and the  on-site Hubbard repulsion $U>0$. Likewise, the  $\mathbb{Z}_2$ fields have an energy difference between the local configurations $\Delta$, and a transverse field of strength $\beta$ that is responsible for their quantum fluctuations. The  $\mathbb{Z}_2$ fields renormalize the bosonic hopping via  $\alpha$, and we set   $\alpha = 0.5t$ and $\Delta = 0.85t$.

This model~\eqref{eq:z2bhm} hosts Peierls-type phenomena  analogous to the fermionic SSH model~\cite{ssh_polymers} but, remarkably, in absence  of a Fermi surface~\cite{z2bhm_BOW_phases}. There are, however, important differences: at half filling and in the  slow-lattice limit  relevant for polyacetylene~\cite{POUGET2016332}, a Peierls' instability inevitably occurs for arbitrarily-small fermion-lattice couplings~\cite{peierls}. In this limit,   the fermionic  groundstate in one of SSB sectors is adiabatically connected  to a free-fermion SPT phase~\cite{classification_review}. In contrast, our bosonic Peierls phases allow for a  genuinely-correlated  topological  bond-ordered wave (TBOW$_{1/2}$) protected by bond-inversion symmetry, which  cannot be adiabatically connected to a free-boson SPT phase~\cite{z2_bhm_SPT_phases}. In this case, the symmetry protecting the TBOW$_{1/2}$ is completely fixed from the dimerization SSB pattern of  the $\mathbb{Z}_2$ fields. 

We now describe a richer situation at filling $n=2/3$ (similarly $n = 1/3$). Here, the Peierls-type SSB leads to a trimerization of the $\mathbb{Z}_2$ field, namely a periodic repetition of a 3-site unit cell with  bonds characterized by arbitrary expectation values  $\langle\sigma^z_{1,2}\rangle,\langle\sigma^z_{2,3}\rangle,\langle\sigma^z_{3,4}\rangle$. 
Note that this trimerization still leaves freedom for various bond configurations that do not necessarily imply a protecting symmetry for the bosons (Fig.~\ref{fig:emergence}(b)). One of the main results of our work is to show how such a protecting symmetry actually emerges dynamically  at low energies, giving rise to  an intertwined TBOW$_{2/3}$ (Fig.~\ref{fig:emergence}(d)). 

We first study a  system of $L = 30$ sites with  DMRG~\cite{tenpy}, for $\beta = 0.01\,t$ and different Hubbard interactions $U$. For weak interactions, the $\mathbb{Z}_2$ field is polarized along the same axis (Fig.~\ref{fig:emergence}(a)), and the bosons display a quasi-superfluid behaviour.  Increasing the interactions leads to a bosonic Peierls transition, whereby  translational symmetry  is spontaneously broken, leading to a three-fold degenerate ground-state with ferrimagnetic-type ordering  $\langle\sigma^z_{1,2}\rangle=\langle\sigma^z_{2,3}\rangle>\langle\sigma^z_{3,4}\rangle$, together with  a bosonic period-3 BOW that displays inversion symmetry with respect to the central inter-cell bond (see Fig.~\ref{fig:berry_trimer}(a,b), left panel). This phase exhibits similar properties to the charge density waves in extended Hubbard models~\cite{Guo2012,Budich2013}, albeit without the need of longer-range interactions.  We  note that a fermionic counterpart of this phase has been predicted  in charge-transfer  salts~\cite{fractionalization_trimer,PhysRevB.27.370}. To characterize its topology, we use the local Berry phase $\gamma^\mu=\ii\int_{0}^{2\pi}{\rm d}\theta\langle\Psi^\mu(\theta)|\partial_\theta\psi^\mu(\theta)\rangle$, where $\ket{\psi^\mu_{\theta}}$ is the $\mu$-th groundstate of the Hamiltonian~\eqref{eq:z2bhm} with a single       bond twisted according to $t\to t\ee^{\ii\theta}$~\cite{hatsugai_local}. The left panel of Fig.~\ref{fig:berry_trimer}(c)  depicts the local Berry phase for one of the  groundstates, which  clearly  vanishes on the inter-cell bonds relevant for the  inversion symmetry of Fig.~\ref{fig:emergence}.  We note that the three possible groundstates become degenerate in the thermodynamic limit,  which can  be characterized by the total Berry phase $\gamma=\sum_{\mu}\gamma^\mu$. For the present  BOW$_{2/3}$, we  also find  $\gamma=0$, indicating  that this  phase is topologically trivial.
  
By further increasing the  interactions,  a phase with a different SSB pattern $\langle\sigma^z_{1,2}\rangle=\langle\sigma^z_{2,3}\rangle<\langle\sigma^z_{3,4}\rangle$ arises (right panels  Fig.~\ref{fig:berry_trimer}(a,b)). Although the ferrimagnetic and BOW patterns look rather  similar to the previous case, the local Berry phase at the inter-cell bonds is  now quantized to   $\gamma_\mu=\pi$ (right panel Fig. \ref{fig:berry_trimer}(c)). Note again that this phase presents  three degenerate groundstates   in the thermodynamic limit,  and we find a total Berry phase  $\gamma=\pi$, indicating a non-trivial  TBOW$_{2/3}$ phase. This exemplifies the scenario of Fig.~\ref{fig:emergence}: from all the  trimerized configurations possible a priori, the system chooses the one with additional bond-centered inversion symmetry, allowing for a topological crystalline insulator~\cite{classification_review}.  In combination with the local order parameters (right panel Fig.~\ref{fig:berry_trimer}(a,b)), this shows that the TBOW$_{2/3}$ is an interaction-induced intertwined topological phase in which, contrary to  half-filling~\cite{z2bhm_BOW_phases}, the protecting symmetry is emergent and not fixed a priori by the SSB pattern.  The ocurrence of this mechanism is a hallmark of our $\mathbb{Z}_2$-Bose-Hubbard model and does not have an analogue  in the  standard SSH model~\cite{fractionalization_trimer,PhysRevB.27.370}.

\begin{figure}[t]
  \includegraphics[width=1.0\linewidth]{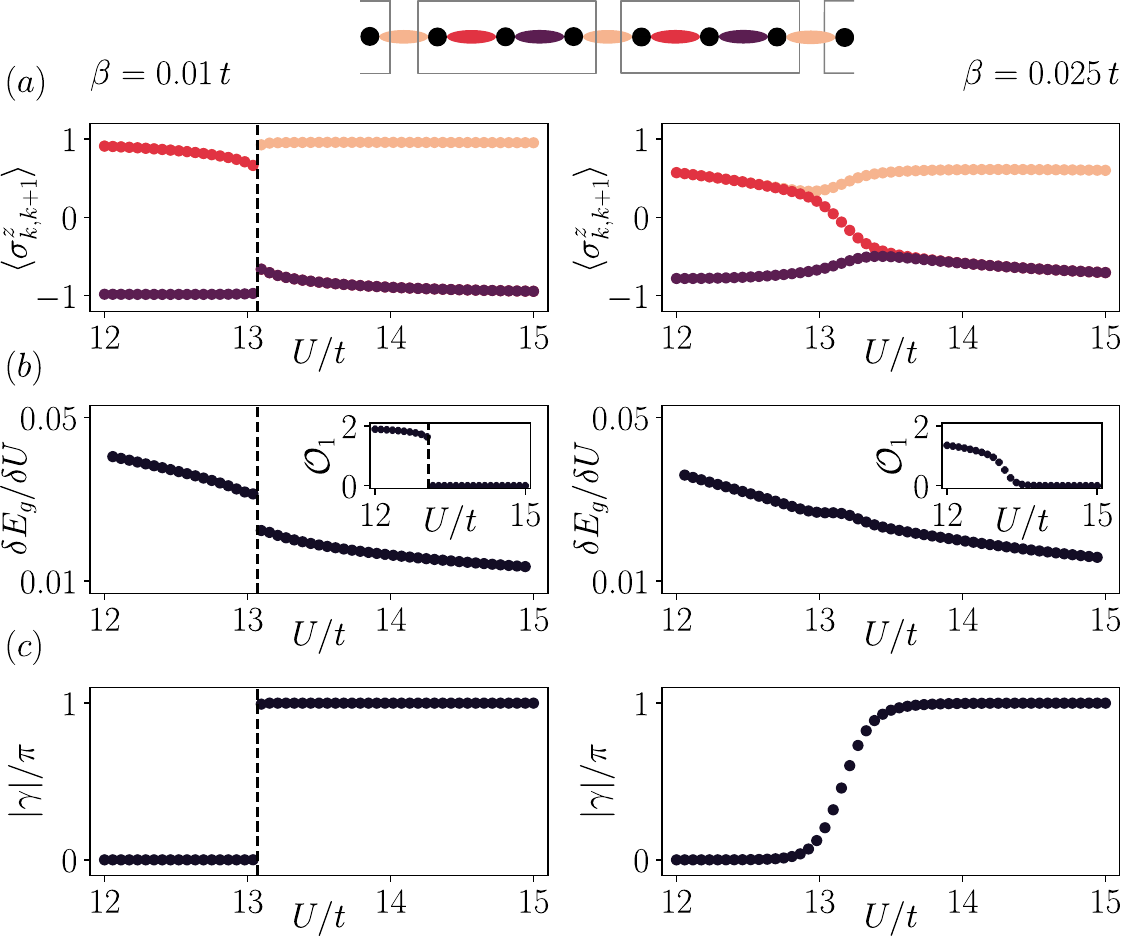}
\caption{\label{fig:trimer_transition}\textbf{Interaction-induced topological phase transitions:} (a) Unit-cell fields $\langle \sigma^z_{k,k+1} \rangle$ as we increase $U$, where $k \in \{1, 2, 3\}$ are the three different bonds. The left and right panels correspond to $\beta = 0.01t$ and $\beta = 0.025t$, respectively. In the first case, there is an abrupt transition between the trivial and topological BOW. In the second case, the transition is continuous, and we find  a finite region where inversion symmetry is broken. (b) First derivative of the ground-state energy $E_{\rm g}$ through the transition. For $\beta = 0.01t$ there is a discontinuous jump, signalling a first-order topological phase transition. The inset shows also a jump in the observable $\mathcal{O}_1$. For $\beta = 0.025t$, both quantities behave smoothly. (c) Total berry phase, where the same behaviour is observed.}
\end{figure}

\paragraph{Interaction-induced topological phase transitions.-- } Topological phase transitions delimiting free-fermion SPT phases, and those found due to their competition with SSB phases, are  typically continuous second-order phase transitions. In the presence of strong correlations, however, first-order topological phase transitions may arise~\cite{first_order_topo_1,first_order_topo_2}. We now discuss how critical lines of  different orders delimit  the intertwined TBOW$_{2/3}$ in a strongly-interacting  region of parameter space, showing that the TBOW$_{2/3}$  cannot be adiabatically connected to a free-boson SPT phase. 

In the adiabatic regime $\beta\ll t$, we observe that the transition between  trivial BOW$_{2/3}$ and intertwined  TBOW$_{2/3}$ is of  first order using an infinite DMRG algorithm~\cite{tenpy}. Figure~\ref{fig:trimer_transition}(a) shows the Ising fields $\langle \sigma^z_{k,k+1} \rangle$ within the unit cell as  the Hubbard interaction is increased, while keeping $\beta$ fixed. For $\beta = 0.01t$~(left column) we observe an abrupt  transition characterized by a discontinuity in the first derivative of the ground state energy 
${\delta E_{\rm g}}/{\delta U} = ({E_{\rm g}(U + \Delta U) - E_{\rm g}(U)})/{\Delta U}$ \cite{first_order_topo_2}, signaling  a first-order phase transition (Fig.~\ref{fig:trimer_transition}(b)). Introducing  the  bond observables,
$
\mathcal{O}_{k} = \bra{E_{\rm g}} \sigma^z_{k, k+1} - \sigma^z_{k+1, k+2} \ket{E_{\rm g}}
$
with $k$ even or odd, we can characterise the corresponding bond-inversion symmetry within the unit cell.  The inset of Fig.~\ref{fig:trimer_transition}(b) shows how $\mathcal{O}_1$ displays a discontinuous jump.
We note that the gap of the system remains open during this transition, while the total Berry phase, computed here with the help of  the entanglement spectrum \cite{ent_spectrum_berry}, changes abruptly, as depicted in Fig.~\ref{fig:trimer_transition}(c). To the best of our knowledge, this is the first topological characterization  of  a first-order  phase transition in an intertwined topological phase.

The situation changes as one departs from the adiabatic regime. Figure~\ref{fig:trimer_transition} (right panel) shows a continuous second-order transition both in $\delta E_{\rm g} / \delta U$ and in $\mathcal{O}_1$ for $\beta = 0.025t$. Remarkably, there is a finite region between the trivial and topological BOW phases where the $\mathbb{Z}_2$ fields have different expectation values, breaking the emergent inversion symmetry within the larger Peierls' trimerization. These results are in accordance with the behaviour of the total Berry phase in Fig.~~\ref{fig:trimer_transition}(c), which shows a non-quantized value in this intermediate asymmetrical  region. In fact, the appearance of this region originates from a very interesting interplay between the emergent inversion symmetry and the Peierls SSB phenomenon: a direct continuous transition between the trivial and topological BOWs would require a gap closing point in the bosonic sector, where every bond had the same expectation value and the BOW would disappear. However, this  comes with an energy penalty, since the Peierls' mechanism favors the formation of a 3-site unit cell~\cite{z2bhm_BOW_phases}. Therefore, the system  energetically prefers to keep the trimerized unit cell at the expense of  breaking the bond-inversion symmetry within the unit cell, and continuously setting the emergent inversion symmetry responsible for the quantized Berry phase $\gamma=\pi$ of  Fig.~~\ref{fig:trimer_transition}(c).  This  non-trivial interplay between symmetry protection and symmetry breaking, driven solely by  correlations, is another hallmark of our $\mathbb{Z}_2$-Bose-Hubbard model, absent at other fillings or in the fermionic SSH model~\cite{fractionalization_trimer,PhysRevB.27.370}. 

\begin{figure}[t]
	\includegraphics[width=1.0\linewidth]{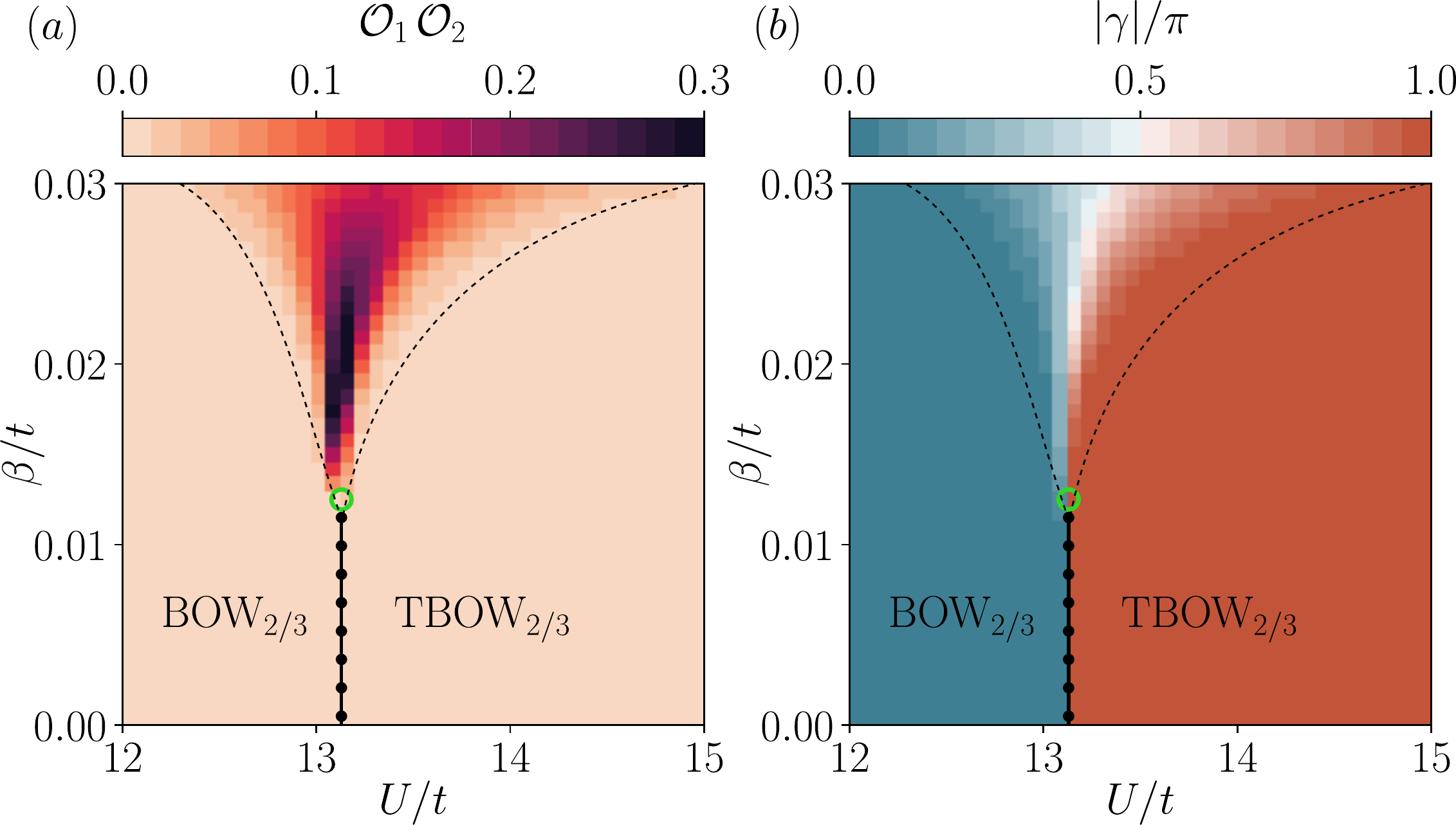}
	\caption{\label{fig:phase_diagram}\textbf{Phase diagram: } (a) In the background, we represent the product of observables   $\mathcal{O}_1\mathcal{O}_2$, which has a non-zero value only in the intermediate phase where bond-inversion symmetry is broken. The solid line marks the first-order critical points separating BOW$_{2/3}$ and TBOW$_{2/3}$ phases. The line ends at a triple point (green circle), after which  a finite region (qualitatively comprised between the dotted lines), where the  bond-inversion symmetry is broken, interpolates between the trivial and topological BOWs. (b) We also present the total berry phase. The latter has a non-quantized value in the region where the protecting inversion symmetry is broken.}
\end{figure}

Finally, we present the phase diagram as a function of $\beta$ and $U$ in Fig.~\ref{fig:phase_diagram}(a) by depicting the product of $\mathcal{O}_1 \mathcal{O}_2$: it can only attain a non-zero value if the bond inversion symmetry within the unit cell is broken  (i.e. if the transition occurs continuously via an intermediate non-symmetric region). As can be neatly observed in this figure, for $\beta < \beta_c$, a first-order transition separates the trivial and topological BOWs.  We find that  $\beta_{\rm c} \approx 0.012 t$, which is consistent with a triple point separating the first-order and continuous transitions that delimit the TBOW$_{2/3}$ (Supplementary Material). Figure.~\ref{fig:phase_diagram}(b) shows the phase diagram in terms of the total Berry phase, quantized to $0$ and $\pi$ in the regions with inversion symmetry and with non quantized values in the region where the symmetry is broken.

\begin{figure*}[t]
\centering
	\subfigure[]{
	\includegraphics[width=0.32\linewidth]{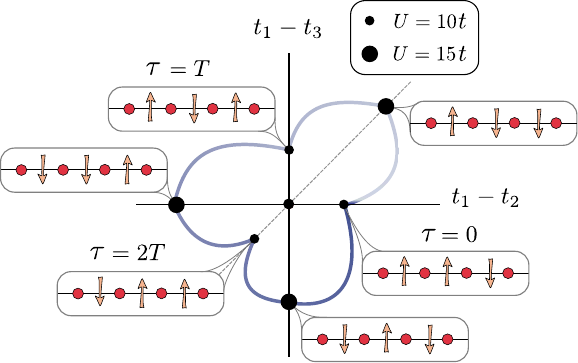}
	}
	\subfigure[]{
	\includegraphics[width=0.32\linewidth]{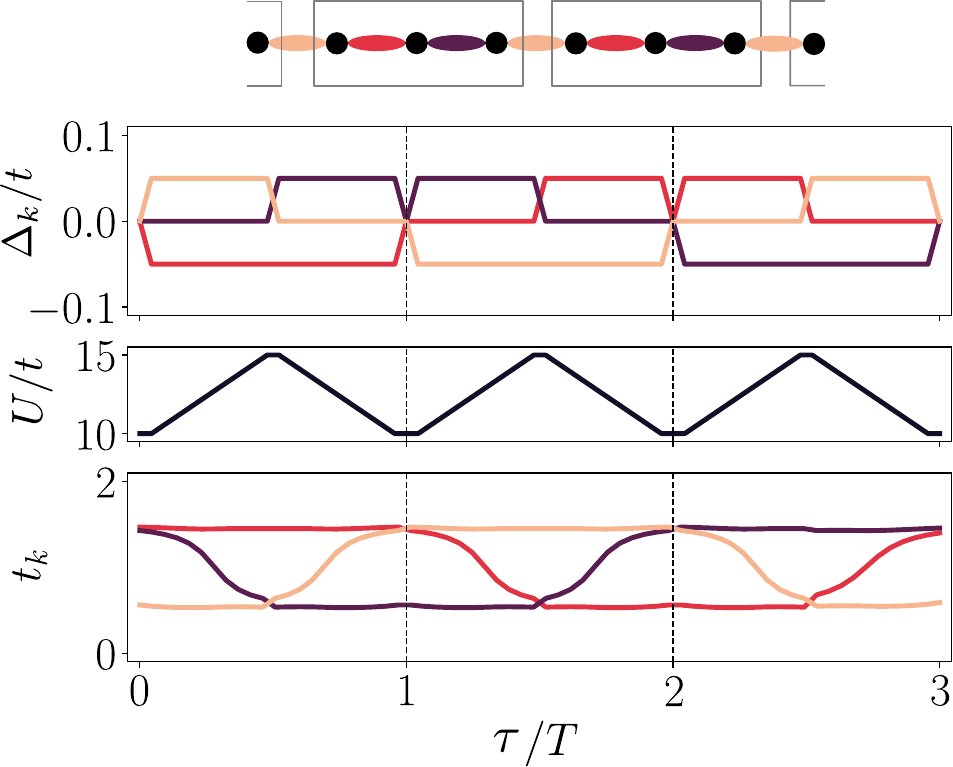}
	}
	\subfigure[]{
	\includegraphics[width=0.32\linewidth]{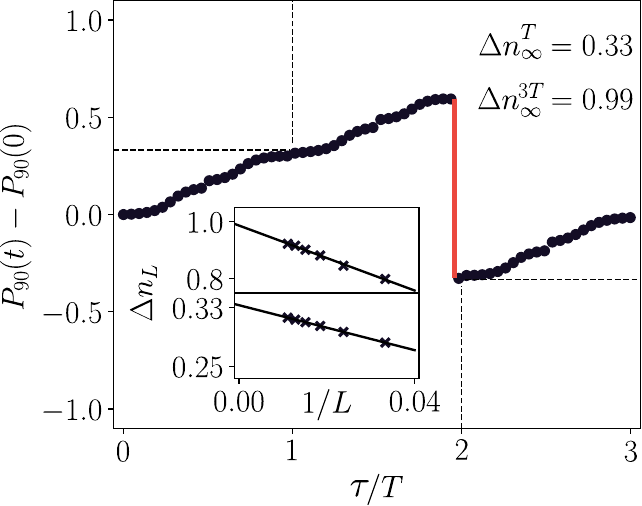}
	}
\caption{\label{fig:pumping_interactions}\textbf{Self-adjusted fractional pumping: }
(a) The trivial and topological BOW phases are three-fold degenerate each. The six different states are represented here as different points on an effective parameter space characterized by the expectation values of the bond fields  $t_k = 1 + \alpha/t \langle{\sigma}_{k,k+1}^z\rangle$, with $k \in \{1, \, 2, \, 3\}$. To define an adiabatic cycle  through these different BOWs, the protecting inversion symmetry must be broken at intermediate states in order to enclose the degeneracy point at $t_1 = t_2 = t_3$. (b) The Peierls mechanism forces the system to break this symmetry spontaneously when interactions are increased, connecting states in the trivial and topological BOW phases (Fig.~\ref{fig:trimer_transition}(b)). In order to select which state from the degenerate manifold the system will transition to, we introduce an external inhomogeneous $\mathbb{Z}_2$ field $\Delta_k$ that is only applied to a subset bonds within the unit cell. The fields partially break the degeneracy  of the BOWs, and restrict the possible adiabatic evolution. A sequential combination of local fields  and interaction-driven self-adjustments allows the system to cycle around the degeneracy point in the effective parameter space. Note that the protocol must be repeated three times for the ground state to reach the initial configuration. (c) COM $P_L(t)$ through the cycle for a finite chain of size $L=90$ and for $\beta = 0.025t$. The discontinuous jump (red), related to the presence of edge states, allows us to obtain the total charge transported in the bulk during one cycle, $\Delta n_{L=90} = 0.92$. Inset: finite-size scaling yields a transported fractional charge at $\tau=T$, and an integer charge at the end of the adiabatic path ($\tau=3T$)}
\end{figure*}

\paragraph{Self-adjusted Fractional pump.-- } Topology can also become manifest through dynamical effects, such as the  quantized transport of charge in electronic systems  evolving under  cyclic adiabatic modulations: Thouless pumping~\cite{pumping_thouless}. This topological pumping lies at the heart of our current understanding of free-fermion SPT phases~\cite{ti_review_2}, and can also be generalized to weakly-interacting systems~\cite{pumping_thouless_int}. As advanced in the introduction, 1D and quasi-1D systems at sufficiently-strong interactions  can exhibit a new type of fractional pumping~\cite{grudst2014,Grushin2015,Zeng2015,Zeng2016,li2017,Taddia2017} that cannot be accounted for using non-interacting topological pumping.

In this section, we show that adiabatic dynamics traversing through intertwined topological phases allows for a new effect: a self-adjusted  fractional pumping   due to the interplay  of the SSB mechanism and other gap-opening perturbations. By introducing  guiding fields that only act on a subset of the $\mathbb{Z}_2$ fields, and  raising/lowering the Hubbard interactions, the  free $\mathbb{Z}_2$ fields  self-adjust dynamically during the adiabatic cycle. As a consequence, the bosonic sector traverses a sequence of groundstates that are energetically favourable due to the Peierls' mechanism. In this way, the system self-adjusts along this adiabatic sequence, allowing for an exotic fractional pumping induced by interactions \cite{grudst2014,Grushin2015,Zeng2015,Zeng2016,li2017,Taddia2017}. The details of this self-adjusted topological pumping are explained in Figure~\ref{fig:pumping_interactions}. 

For finite systems, the pumped charge can be inferred from the center of mass (COM)
$P_L(\tau) = \frac{1}{L} \sum_j (j-j_0) \bra{\Psi(\tau)} \hat{n}_j \ket{\Psi(\tau)}$, 
where $j_0$ is the center of a chain of size $L$, and $\ket{\psi(\tau)}$ is the adiabatically-evolved state  at time $\tau$. Figure~\ref{fig:pumping_interactions}(c) shows the DMRG results describing how the COM changes along the cycle connecting the  BOW$_{2/3}$ and TBOW$_{2/3}$ possible groundstates for a finite chain of size $L=90$. After $\tau=T$, we observe a COM displacement of $\Delta n^{T}_{L=90} = P_{L=90}(T) - P_{L=90}(0) = 0.316$, reflecting the fractional charge. To obtain precisely the charge, we perform a finite-size scaling analysis and find $\Delta n^{T}_{\infty} = \lim_{L\to\infty} \Delta n^{T}_{L} = 1/3$ (inset). At $\tau=2T$, the COM displacement reaches a value consistent with $2/3$ in the thermodynamic limit. We note that these fractional values are characteristic of a strongly-correlated SPT phase with groundstate degeneracy, and cannot be found for any non-interacting topological phase (Supplementary Material). In our present case, the adiabatic path in parameter space can be understood as a dynamical analogue of the spatial interpolation between the different groundstates, which leads to topological solitons and fractionally-quantized charges bound to them~\cite{fractionalization_trimer}. During each period $T$, we  interpolate between two such groundstates, and a fractional charge is pumped   without  creating any spatial solitonic profile. 

Let us now turn our attention to the discontinuous jump of the pumped charge towards $-1/3$, as this is related to the presence of many-body edge states for a finite system~\cite{pumping_bulk_edge}, and can be used to define a bulk-boundary correspondence for our intertwined TBOW$_{2/3}$. The transported charge across  the bulk, $\Delta n^{3T}_L$, can be related to the discontinuous jumps during the cycle~\cite{pumping_bulk_edge}, namely
$
\Delta n^{3T}_L = - \sum_i \Delta P_L(\tau_i),$ where $ \Delta P_L(\tau_i) = P_L(\tau_i^+) - P_L(\tau_i^-)
$ quantify the discontinuities occurring at instants 
 $\tau_i$, and $\tau_i^\pm = \tau_i \pm \epsilon$ with $\epsilon\to0$. In the thermodynamic limit, it converges to the quantized value of the pumped charge $\Delta n^{3T}_{\infty} = \lim_{L \to \infty} \Delta n^{3T}_L = 1$  related to the integer Chern number in an extended 2D system~\cite{pumping_bulk_edge}. Since these discontinuities depend on the presence of edge states in a finite system, the center-of-mass approach establishes a sort of bulk-boundary correspondence that can be explicitly proven via the adiabatic pumping. Moreover, the COM can be measured in cold-atomic experiments \cite{pumping_troyer}, and it has been used to reveal the topological properties of fermionic and bosonic SPT phases~\cite{pumping_exp_fermions, pumping_exp_bosons}.

By estimating the discontinuity, we can extract the transported charge across  the bulk during the whole adiabatic evolution that bring the BOW back to itself after $\tau=3T$, obtaining a nearly quantized value $\Delta n_{L=90} = 0.92$. As it is shown in the inset, a truly quantized charge is recovered in the thermodynamic limit, signaling the topological nature of the system. These results allow us to establish a bulk-boundary correspondence in the pumping process \cite{pumping_bulk_edge}, even though this was not guaranteed a priori due to the lack of the global  symmetries regarding the ten-fold classification of topological insulators. In particular, one may understand the edge states of the TBOW$_{2/3}$ as remains of topologically-protected conducting edge states of an extended 2D system (Supplementary Material). We note that, even if the topological degeneracy point does not appear in the phase diagram of the model, the quantized transported charge reveals its presence in an effective parameter space, as a non-zero quantized charge can only be obtained when the parameter modulation encircles such a degeneracy point.

\paragraph{Conclusions and outlook.--} We have shown how symmetry protection can emerge dynamically in an interplay between symmetry breaking and strong correlations. In the $\mathbb{Z}_2$ Bose-Hubbard model, this novel mechanism gives rise to intertwined topological phase for certain fractional fillings. The unique properties of these phases are manifest in the special static and dynamical features discussed in this work. A realistic implementation of the model with cold atoms is suggested by recent experimental results \cite{z2LGT_1, z2LGT_2}. The proposed self-adjusted pumping protocol, in particular, could be used to reveal the topological properties of the system and its fractional nature. Future research directions include the study of topological defects on top of the intertwined topological phases, where localized states with fractional particle number are expected to appear, signaling deeper connections to the physics of the FQHE.

\paragraph{Methods. --} The numerical calculations have been performed using a density matrix renormalization group algorithm (DMRG) \cite{tenpy}. For the finite-size calculations we used a matrix product state (MPS) based algorithm with bond dimension $D = 100$. To directly access the thermodynamic limit we used an infinite MPS with a repeating unit cell composed of three sites and $D = 150$. The Hilbert space of the bosons is truncated to a maximum number of bosons per site of $n_0 = 2$. This is justified for low densities and strong interactions.

\paragraph{Supplementary Information.--}See Supplementary Material for further details on the topological phase transitions and the location of the tricritical point, as well as on the connection between the Thouless pumping and an extended 2D topological system, shedding light on the nature of fractionalization and the presence of edge states.

\paragraph{Acknowledgements.--}The authors thank L. Tagliacozzo for useful discussions. This project has received funding from the European Union's Horizon 2020 research and innovation programme under the Marie Sk\l{}odowska-Curie grant agreement No 665884, the Spanish Ministry MINECO (National Plan 15 Grant: FISICATEAMO No. FIS2016-79508-P, SEVERO OCHOA No. SEV-2015-0522, FPI), European Social Fund, Fundaci\'{o} Cellex, Generalitat de Catalunya (AGAUR Grant No. 2017 SGR 1341 and CERCA/Program), ERC AdG OSYRIS, EU FETPRO QUIC, and the National Science Centre, PolandSymfonia
Grant No. 2016/20/W/ST4/00314. A. D. is financed by a Juan de la Cierva fellowship (IJCI-2017-33180). A.B. acknowledges support from the Ram\'on y Cajal program  RYC-2016-20066,  MINECO project FIS2015-70856-P, and   CAM/FEDER Project  S2018/TCS-4342 (QUITEMAD-CM).

\bibliographystyle{apsrev4-1}
\bibliography{bibliography}

\end{document}



\title{ Supplemental Material: Intertwined Topological Phases induced by Emergent Symmetry Protection}

\begin{abstract}

In this Supplementary Material we discuss additional details concerning the topological phase transitions in the $\mathbb{Z}_2$ Bose-Hubbard model (1) and the location of the triple point, as well as on the connection of the 1D model to an extended 2D topological system, shedding light on the nature of fractionalization and the presence of edge states.

\end{abstract}

\maketitle

\section{Topological Phase Transitions}

In this part of the Supplemental Material, we provide further details concerning the interaction-induced topological phase transitions discussed in the main text. We recall that the parameters $\Delta = 0.85\,t$ and $\alpha = 0.5\,t$ are fixed henceforth.

\subsection{Gap closure and interaction-induced nature of TBOW$_{2/3}$}

The principle of adiabatic correspondence, which states that groundstates that can be adiabatically deformed into each other belong to the same phase of matter, can be used as a primary tool to group groundstates according to a common set of underlying properties. This principle lies at the heart of the modern definition of topological insulators as insulating phases of matter characterized by a topological invariant that cannot be adiabatically connected to the atomic limit (i.e. trivial band insulator). In the context of SPT phases, this principle can be used to delimit the region in parameter space hosting a correlated topological phase, as it must be separated from other states of matter via an abrupt phase transition (i.e. first-order discontinuous transition, or second-order phase transition marked by a gap closure) where the adiabaticity is lost.

In the present context, we have clearly shown in the main text that the TBOW$_{2/3}$ can be connected to the trivial BOW$_{2/3}$ via a discontinuous first-order transition. Additionally, we mentioned that for stronger transverse fields, this transition becomes a continuous second order transition that separates these phases from an intermediate region where inversion symmetry is broken. It is interesting to note that, although the groundstate of the system preserves a trimerized unit cell to satisfy the underlying Peierls mechanism and minimize its energy, the gap of the system does indeed close in this intermediate region, which  guarantees that the TBOW$_{2/3}$ is a  well-defined phase that cannot be adiabatically connected to   the trivial BOW$_{2/3}$. In order to show the occurrence of such a gap closure,  we use  entanglement spectroscopy in the thermodynamic limit. 
 
\begin{figure}[t]
  \centering
  \includegraphics[width=1.0\linewidth]{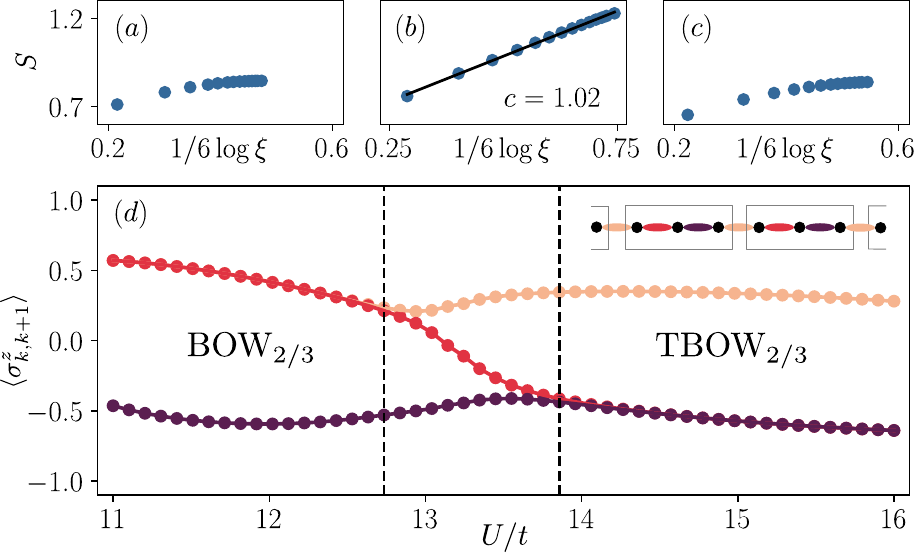}
\caption{\label{fig:entropy_scaling}\textbf{Entropy scaling: }From (a) to (c) we show the entanglement entropy $S$ for an iMPS as a function of the correlation length $\xi$ for a state in the BOW$_{2/3}$, intermediate region and TBOW$_{2/3}$, respectively. In the BOW phases, the entropy saturates, signaling a gapped ground state. In the intermediate region, the scaling is logarithmic. Different points are calculated using different bond dimensions $D$, and we set $\beta = 0.03t$. (d) Evolution of the $\mathbb{Z}_2$ field on the bonds of the unit cell $\langle \sigma^z_{k,k+1}\rangle$, with $k \in \{1, 2, 3\}$ in terms of $U$. The three different phases are separated qualitatively by two dotted lines.}
\end{figure}

Figure~\ref{fig:entropy_scaling} shows the scaling of the entanglement entropy $S(\rho_{\ell})=-{\rm Tr}\{\rho_{\ell}\log(\rho_{\ell})\}$, where $\rho_{\ell}$ is the reduced  density matrix for a bipartition of the groundstate into two blocks. This entanglement entropy is expressed  in terms of the infinite matrix-product state (iMPS) correlation function $\xi$ for different values of the bond dimension $D$ and for different Hubbard interactions $U$, where we set $\beta = 0.03t$. If the ground state is gapped, this entropy  saturates~\cite{finite_D_scaling}. On the contrary, in a gapless critical point/region, the following scaling relation holds 
\begin{equation}
S = \frac{c}{6}\,\text{log}\,\xi,
\end{equation}
for sufficiently large $\xi$. To capture this saturation or the logarithmic scaling, instead of performing a finite-size scaling, one may conduct a  finite-$D$ scaling, which shows how both BOW$_{2/3}$ (Fig.~\ref{fig:entropy_scaling}(a)) and TBOW$_{2/3}$ are gapped (Fig.~\ref{fig:entropy_scaling}(c)). In the intermediate symmetry-broken region, we find instead a logarithmic scaling of the entanglement entropy consistent with a conformal charge $c=1$ (Fig.~\ref{fig:entropy_scaling}(b)). Accordingly, we can conclude that both phases cannot be adiabatically connected, neither at weak nor at stronger transverse fields. In the regime of small quantum fluctuations (i.e. weak transverse fields),  a first-order phase transition takes place. For larger quantum fluctuations (i.e. stronger transverse fields), the gap closes and a a second-order phase transition occurs via a critical region where the emergent inversion symmetry is absent. We note that, when the guiding fields are added in the pumping protocol, the total gap of the system remains open during the  cycle, guaranteeing a well-defined adiabatic pumping.

\subsection{Triple point separating first- and second-order transitions}

\begin{figure}[t]
  \centering
  \includegraphics[width=1.0\linewidth]{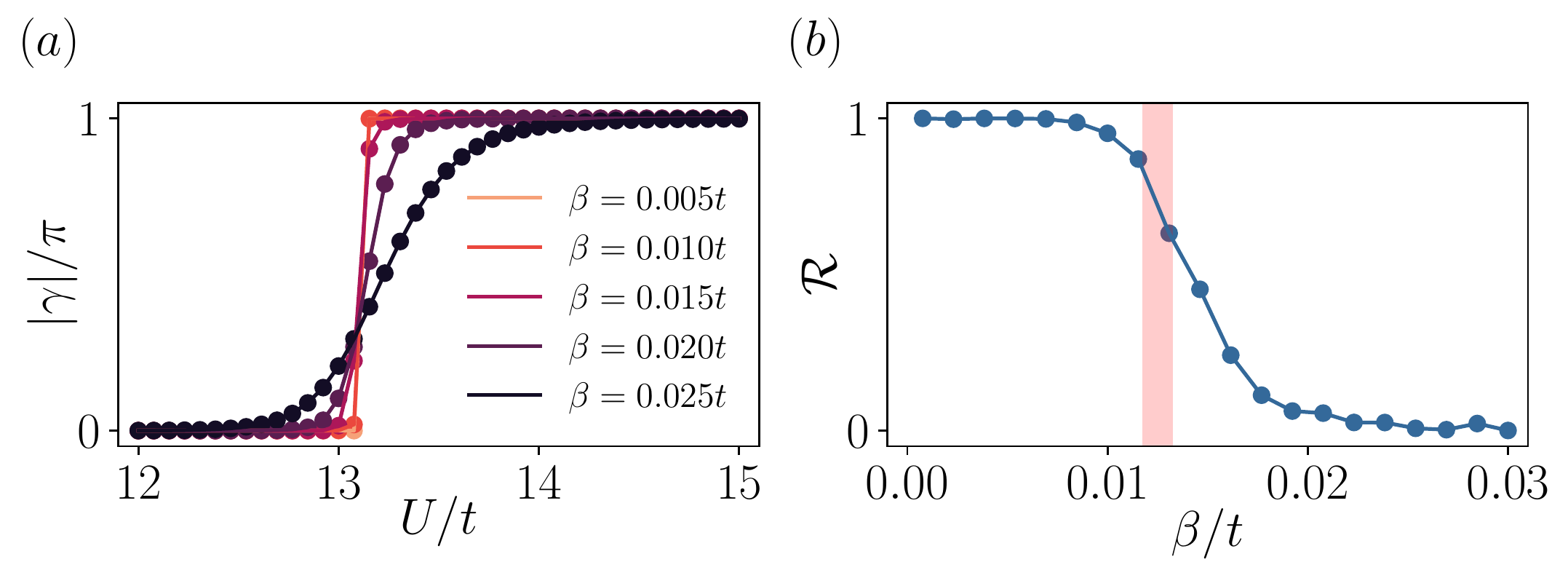}
\caption{\label{fig:triple}\textbf{Triple point: }(a) Change in the topological invariant $\gamma$ through the interaction-induced topological phase transition for different values of $\beta$. (b) Value of $\mathcal{R}$ (see text) in terms of $\beta$, characterizing the type of phase transition. At $\beta_c \approx 0.012\,t$, it changes from a first-order to a continuous transition.}
\end{figure}

As noted above and in the main text,  the topological phase transition between the trivial and topological BOW phases changes from first-order to continuous at some critical value $\beta_c$. In Figure~\ref{fig:triple}(a) we can see how the topological Berry phase $\gamma$, as defined in the main text, changes as we increase interactions for different values of $\beta$. In this part of the Supplemental Material, we show how the value of $\beta_c$ can be found  by analyzing the first derivate of the Berry phase
\begin{equation}
\frac{\delta \gamma}{\delta U} (U, \Delta U) = \frac{\gamma(U + \Delta U) - \gamma(U)}{\Delta U}
\end{equation}
In particular, we calculate this quantity for two different discretizations, $\Delta U$ and $2 \Delta U$. For a first-order  transition, $\gamma$ shows a discontinuous jump at the transition point $U_c$, and the finite differences $\frac{\delta \gamma }{ \delta U} (U_c, 2\Delta U) \rightarrow \half\frac{\delta \gamma }{ \delta U} (U_c, \Delta U)$ for $\Delta U \rightarrow 0$. In contrast, for a continuous second-order transition, one finds that  $\frac{\delta \gamma }{ \delta U} (U_c, 2\Delta U) \rightarrow \frac{\delta \gamma }{ \delta U} (U_c, \Delta U)$. Accordingly, the  quantity
\begin{equation}
\mathcal{R} = \frac{|\frac{\delta \gamma}{\delta U} (U_c, \Delta U)|  - |\frac{2\delta \gamma}{\delta U} (U_c, 2\Delta U) |}{|\frac{\delta \gamma}{\delta U} (U_c, 2\Delta U) |}
\end{equation}
must converge to $1$ or $0$ for first-order and continuous transitions, respectively \cite{first_order_topo_2}.

 In Fig.~\ref{fig:triple}(b) we represent this quantity for different values of $\beta$ for $\Delta U = 0.033t$, which clearly show this different qualitative scalings. The results are consistent with a critical value of $\beta_c \approx 0.012t$, corresponding to a triple point separating the first- and second-order transitions.

\section{Absence of fractionalization in the effective  single-particle pumping}

In this part of the Supplemental Material, we clarify different aspects of the pumping scheme introduced in the main text using an effective non-interacting model. We show that, whereas some qualitative features of the topological pumping in the $\mathbb{Z}_2$ Bose-Hubbard model can be understood through the non-interacting analogue, the very fractional nature lacks a non-interacting simile, and must be consider as a direct manifestation of the strongly-correlated nature of the TBOW$_{2/3}$.

\subsection{Bulk-edge correspondence and Chern numbers}

Consider the following non-interacting Hamiltonian,
\begin{equation}
\label{eq:eff_ham}
\hat{H}_\mathrm{eff} = -\sum_i t_i \left(\hat{c}^\dagger_i\hat{c}^{\vphantom{\dagger}}_{i+1}+\text{H.c.}\right)
\end{equation}
where $\hat{c}_i$ and $\hat{c}^\dagger_i$ are fermionic operators. $\hat{H}_\mathrm{eff}$ is related to the hard-core boson limit of the $\mathbb{Z}_2$BHM, with $U \rightarrow \infty$, and a totally adiabatic or classical $\mathbb{Z}_2$ fields ($\beta = 0$) that are treated in a mean-field-like manner. Accordingly, one may consider that the expectation values $\langle\hat{\sigma}^z_{i,i+1}\rangle$ behave  as external parameters that can be used to control the effective tunnelling coefficients, $t_i = 1 + \alpha / t \langle \hat{\sigma}_{i,i+1}^z\rangle$. 

\begin{figure}[t]
  \centering
  \includegraphics[width=0.8\linewidth]{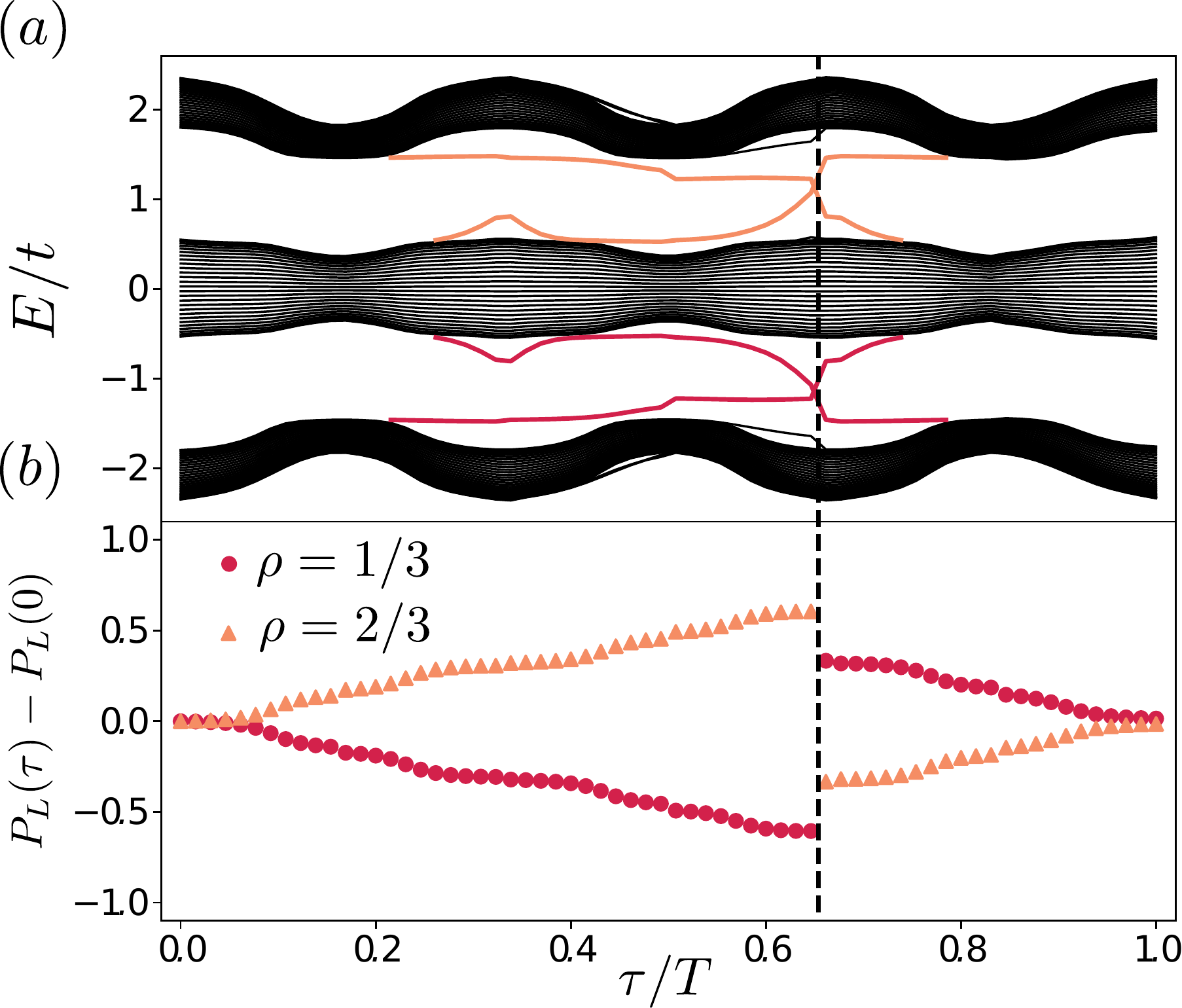}
\caption{\label{fig:pumping_effective}\textbf{Effective pumping: }(a) spectral flow during the adiabatic cycle. At each time $\tau$ we draw the energy levels by diagonalizing the non-interacting effective Hamiltonian (\ref{eq:eff_ham}). We observe three distinct bands that remain open during the cycle. There are two energy levels corresponding to the localized edge states inside each band gap. These states connect the different bands and cannot be adiabatically removed without closing the gap. (b) Change in the center of mass during the cycle for densities $\rho = 1/3$ and $\rho = 2/3$. A discontinuous jump occurs when the edge states in (a) cross in energy.}
\end{figure}

However, we note that in this simplified effective model, the tunnelings $\{t_i\}$ are free model parameters that can be changed adiabatically at will. In particular, the effective  tunnelling strengths are changed following the same path as for the interacting case (Fig.~4(a) in the main text). Note that the expectation values $\langle\hat{\sigma}^z_{i,i+1}\rangle$ extracted from the ground states throughout the self-adjusted many-body pumping are different in general for different unit cells. Being non-interacting, the Hamiltonian (\ref{eq:eff_ham}) can be exactly diagonalized, and the band structure can be obtained at any instant of the adiabatic cycle. This calculation leads to the spectral flow  shown in Figure~\ref{fig:pumping_effective}(a). Throughout the cycle, the instantaneous energy levels can be arranged into   three different bands separated by two gaps that remain finite through the whole adiabatic path. Note also that these spectral bands are connected by two in-gap modes, which correspond to the localized single-particle edge states crossing at some intermediate instant of the cycle. Let us highlight that, in contrast to the adiabatic path of the full interacting model that is composed of three adiabatic cycles, the non-interacting case consists of a single cycle where the effective tunnelings are periodically modulated. 

Fig.~\ref{fig:pumping_effective}(b) shows the change in time of the center or mass (COM) $P_L(\tau)$, as defined in the main text,  for densities $\rho=1/3$ and $\rho=2/3$, i.e. when  the lowest or the two lowest bands are filled, respectively. We consider a finite chain of size $L=90$, and show that in both cases a discontinuous jump occurs when the corresponding edge states cross in energy, where the COM changes by $1$ or $-1$. For $\rho=1/3$, before this instant of time only the left edge state is occupied, while it gets empty and the right edge state gets occupied right after it. This shows the fractionalization of the edge states: if each possesses a particle number of $1/2$, this process changes the COM by $-(-1/2)+(1/2)=1$. For $\rho=2/3$, the opposite process takes place.

As explained in the main text, we can calculate the transported charge in the bulk from the change in the COM at the discontinuous jumps. We obtain $\Delta n_{L = 90}^{1/3} = -0.94$ and $\Delta n_{L = 90}^{1/3} = 0.94$. The fact that these are not totally quantized is due to finite-size effects, and they would converge to strictly quantized values in the thermodynamic limit. We thus see that this effective model allows for a net transport of a single quantum of charge accross the bulk, either from the left edge to the right one, or vice versa.  Note however that the period $T$ of this single-particle cycle corresponds to the three consecutive cycles $3T$ of the many-body pumping described in Fig.~4(a) in the main text. Therefore, the integer nature of the pumped single-particle charge after one cycle is consistent with the integer value of the pumped many-body charge after three cycles described in the main text.

As noted in the main text, the transported charge in an infinite chain over one period, $\Delta n$, gives access to  topological phases in higher dimensions. In this case, the pumped charge can  be related to the Chern number of an extended two-dimensional system, where time is taken as a synthetic dimension~\cite{berry_phase_review}, namely
\begin{equation}
\label{eq:pumped_charge}
\begin{aligned}
\Delta n =& - \sum_n  c_n, & c_n =& \frac{1}{2\pi} \int_0^T \mathrm{d}\tau \, \int_\mathrm{BZ} \mathrm{d}q \, \Omega^n_{q\tau},
\end{aligned}
\end{equation}
where $c_n$ is the Chern number corresponding to the $n$th band, and the sum is taken over completely filled bands. The Chern number is calculated as a surface integral in the parameter space formed by the time $t$ and momentum $q$, which has the shape of a torus, and allows to define a  Berry curvature
\begin{equation}
\Omega^n_{q\tau} = \ii\left(\left\langle\frac{\partial u_n}{\partial q}\right|\left.\frac{\partial u_n}{\partial \tau}\right\rangle - \left\langle{\frac{\partial u_n}{\partial \tau}}\right|\left.{\frac{\partial q}{\partial q}}\right\rangle\right)
\end{equation}
with $\ket{u_n}$ the $n$-th Bloch eigenstate. A similar relation to Eq.~(\ref{eq:pumped_charge}) holds also in the presence of disorder and interactions by introducing  twisting fields~\cite{pumping_thouless_int}.
We can calculate the associated  Chern number for the effective model (\ref{eq:eff_ham}) in the thermodynamic limit using time as a synthetic dimension. Using the efficient numerical method ~\cite{hatsugai_algorithm}, we obtain $c_1 = 1$ and $c_2 = -2$, for the first and second bands, respectively. Using Eq.~\eqref{eq:pumped_charge}, we obtain the respective  charges, $\Delta n^{1/3} = -1$ and $\Delta n^{2/3} = 1$. 

This quantized pumping allows to shed light on the origin of the edge states of the trimerized configuration. In contrast to the dimerized half-filling model, the topological origin of the single-particle edge states in this case is not guaranteed a priori due to the lack of chiral/sub-lattice symmetry~\cite{hatsugai_topological_edge}. However, they can be understood as remains of edge states in the extended two-dimensional system, which are indeed topologically protected even in the absence of chiral symmetry~\cite{hatsugai_chern}. This is clear from the spectral flow represented in Fig.~\ref{fig:pumping_effective}(a), which can be seen as the band structure of a two-dimensional system in a cylindrical geometry \cite{quasicrystals}. There, the edge states connect the bands separated by a gap, and this means that they  cannot disappear under perturbations that do not close the gap. Although this spectral flow can not be computed in the interacting case, we expect the argument to hold based on the quantization of the pumped charge, and the extension of Eq.~(\ref{eq:pumped_charge}) for many-body systems \cite{pumping_thouless_int}. Therefore, the observation of the integer pumped charge in the $\mathbb{Z}_2$-Bose-Hubbard model can be used as a bulk-boundary correspondence that clarifies the topological origin of the many-body edge states.

\subsection{ Fragility of the non-interacting  fractionalized pumping}

\begin{figure}[t]
  \centering
  \includegraphics[width=1.0\linewidth]{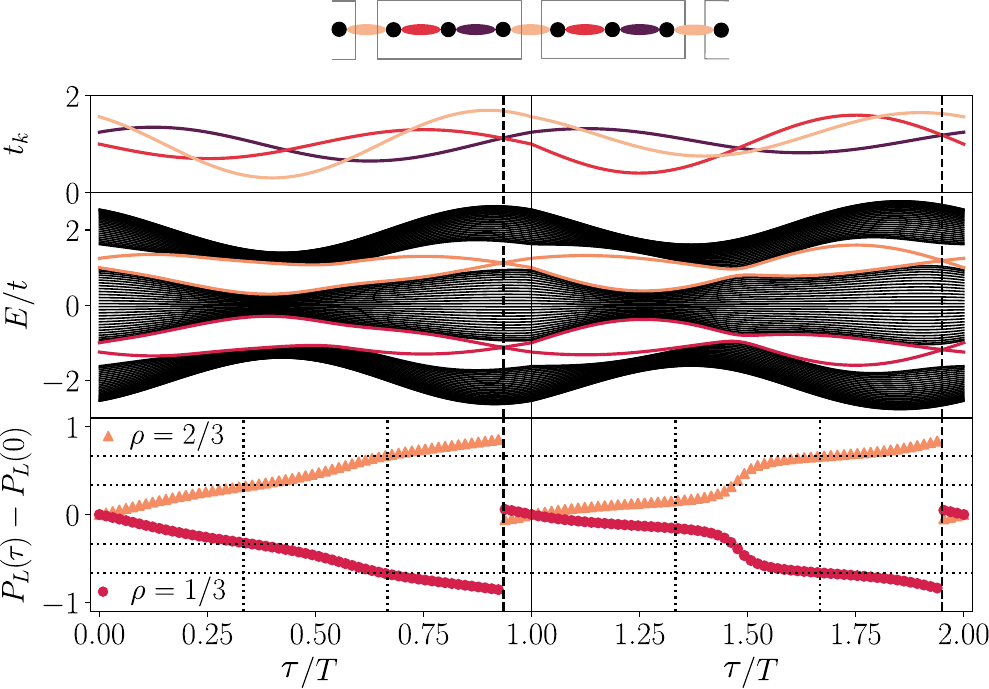}
\caption{\label{fig:fractional_effective}\textbf{Loss of fractionalization: }(a) spectral flow for two different pumping cycles, equal up to adiabatic deformations. (b) Change in the center of mass during the two cycles. Even if for the first case the transported charge at fractions of the cycle might seem quantized, the second case shows that this is not necessary the case.}
\end{figure}

A crucial difference between the many-body pumping presented in the main text, and the effective pumping described in this Supplemental Material is that, for the latter, the robust fractionalization of the pumped charge is absent. The main reason behind this is that the ground state of the effective Hamiltonian (\ref{eq:eff_ham}) is not degenerate in the topological phase. Therefore, the adiabatic path can not be decomposed in three independent periodic cycles as in the protocol presented in the main text. As a consequence the transported charge after a time $T/3$ is not necessarily quantized to $1/3$, only the total charge transported at $T$ is quantized to $1$. This can be seen clearly in Figure~\ref{fig:fractional_effective}. There, we present two adiabatic pumping cycles, connected by a local deformation. Even if in the first case the transported charge might seem to be fractional quantized for fractions of the period, we observe how this fractionalization is lost in the second deformed cycle. Since the latter is just an adiabatic deformation of the first one, we conclude that the fractional charge is not topologically protected. This is different in the many-body pumping presented in the main text: since the total adiabatic path is decomposed in three closed cycles, any adiabatic deformation would preserve the  nature of the pumping, which is robust and quantized to fractional values .

\bibliographystyle{apsrev4-1}
\bibliography{bibliography}